\documentclass[%
reprint,
superscriptaddress,
amsmath,amssymb,
aps,
floatfix,
twocolumn
]{revtex4}
\usepackage{bm}
\usepackage{graphicx}
\usepackage{dcolumn}
\usepackage{bm}
\usepackage{xspace}
\usepackage[]{hyperref}
\usepackage{ulem}
  \hypersetup{      
  unicode=true,         
  pdftoolbar=true,     
  pdfmenubar=true,     
  pdffitwindow=true,    
  pdfstartview={FitH},    
  pdfsubject={Sterilenus@DUNE},  
  pdfnewwindow=true,     
  pdfcreator={RevTeX},
  colorlinks=true,     
  linkcolor=blue,   
  citecolor=blue,    
  filecolor=black,   
  urlcolor=blue,           
  }
\usepackage{placeins}
\usepackage{diagbox}
\usepackage{braket}

\begin{document}

\preprint{APS/123-QED}

\title{Gravitational Wave Signature of Aspherical Bubbles Driven by Thermal Fluctuation}

\affiliation{School of Physics, Henan Normal University, Xinxiang 453007, P. R. China}

\author{Ligong Bian}
\affiliation{Department of Physics and Chongqing Key Laboratory for Strongly Coupled Physics, Chongqing University, Chongqing 401331, P. R. China}

\author{Guangshang Chen}
\email[]{chenguangshang@itp.ac.cn}
\affiliation{Institute of Theoretical Physics, Chinese Academy of Sciences, Beijing 100190, P. R. China}
\affiliation{School of Physical Sciences, University of Chinese Academy of Sciences,  Beijing 100049, P. R. China}


\author{Song Li}
\email[]{chunglee@htu.edu.cn}
\affiliation{School of Physics, Henan Normal University, Xinxiang 453007, P. R. China}

\author{Hongxin Wang}
\affiliation{Schools of Physics, Shandong University, Jinan 250100, P. R. China}

\author{Yang Xiao}
\email[]{xiaoyangphy@gmail.com}
\affiliation{School of Physics, Henan Normal University, Xinxiang 453007, P. R. China}

\author{Jin Min Yang}
\affiliation{School of Physics, Henan Normal University, Xinxiang 453007, P. R. China}

\author{Yang Zhang}
\email[]{zhangyang2025@htu.edu.cn}
\affiliation{School of Physics, Henan Normal University, Xinxiang 453007, P. R. China}

\date{\today}

\begin{abstract}
Cosmological first-order phase transitions are a well-motivated source of stochastic gravitational waves~(GWs), but most predictions are made based on the highly idealized model of perfectly spherical vacuum bubbles, neglecting thermal fluctuations. In this work we use $(3+1)$-dimensional lattice simulations of a scalar model with thermal initial conditions to quantify how thermal fluctuations distort bubble profiles and modify the resulting GW spectrum. We find that thermal fluctuations can strongly break spherical symmetry at early times, allowing even an isolated bubble to emit GWs. In multi-bubble simulations, thermal fluctuations systematically reshape the spectrum, suppressing the infrared part while enhancing and broadening the high-$k$ tail. We further provide an analytical estimate for the ultraviolet regime of the GW spectrum, which is in good agreement with our lattice results and suggests that this regime is dominated by thermal fluctuations. These effects could leave observable imprints in future GW searches.
\end{abstract}

\maketitle

\textbf{\textit{Introduction}}-- First-order phase transitions (FOPTs) in the early Universe provide a powerful probe of fundamental physics, with far-reaching implications for baryogenesis~\cite{Morrissey:2012db,Kuzmin:1985mm, Kawamura_2011}, dark matter production~\cite{Baker:2018vos,Baker:2019ndr,Baker:2017zwx}, primordial black hole formation~\cite{Hawking:1982ga,Carr:2020gox}, and stochastic gravitational waves (GWs)~\cite{Athron:2023xlk}. 
The  transition proceeds through the formation of true-vacuum bubbles within the false vacuum. A widely used starting point for modeling FOPT dynamics is the spherical bubble. The critical bubble profile is described by an exactly $O(4)$-symmetric solution at zero temperature, or an $O(3)$-symmetric one at finite temperature~\cite{Coleman:1977py, Coleman:1977th, Linde:1981zj}. 
This assumption underlies much of the standard picture of nucleation, growth, and collision, and consequently GW predictions.


However, a growing body of theoretical and numerical evidence is beginning to challenge this long-standing assumption. Real-time simulations incorporating thermal fluctuations show that the bubbles formed dynamically in a hot plasma are generically non-spherical, often seeded by oscillons and exhibiting pronounced asphericity during their early growth~\cite{Hertzberg:2020tqa,Bian:2025twi,Batini:2023zpi,Braden:2018tky}. Vacuum decay initiated by cosmic strings can proceed through profiles with an $O(2) \times O(2)$ symmetry rather than the conventional $O(4)$, fundamentally altering the nucleation geometry and action~\cite{Chatrchyan:2025uar}. 
When gravity is taken into account, spherical bubbles can be unstable and can readily develop aspherical distortions~\cite{Aguirre:2005xs}. These developments highlight that the surrounding environment can strongly influence bubble nucleation and growth, motivating a reassessment of FOPT dynamics beyond the idealized symmetric picture.

Bubble nucleation is a fluctuation-driven process via quantum tunneling at $T=0$ or thermal activation at finite temperature, so the ambient stochastic environment is intrinsic to FOPTs. It is therefore natural to expect that fluctuations continue to influence the bubble after nucleation, exciting nonspherical modes and fundamentally altering subsequent evolution and observables. Indeed, quantum fluctuations around an expanding bubble can generate nontrivial multipole moments and source GW emission even from a single bubble~\cite{Blum:2024singlebubble}. Linear stability analyses around the ideal spherical bubble typically suggest that nonspherical perturbations excited near nucleation are rapidly damped as the bubble accelerates~\cite{Adams:1989su}. In contrast, fully nonlinear simulations demonstrate that such perturbations can be amplified rather than suppressed, leading to persistent deformations or even fragmentation into localized oscillons~\cite{Braden:2014cra,Braden:2015vza,Bond:2015zfa}. The evolution of nonspherical fluctuations is therefore governed by nonlinear dynamics, highlighting the necessity of real-time simulations to accurately model bubble evolution and its observable signatures.


In this Letter, we employ (3+1)-dimensional numerical simulations to quantitatively assess the impact of thermal fluctuations on post-nucleation bubble evolution. Our results demonstrate that fluctuations do drive the loss of spherical symmetry during bubble expansion. 
We compute the GW spectrum for both a single aspherical bubble and multi-bubbles,  providing a numerical and a semi-analytic quantification of the influence of thermal fluctuations.

\vspace{.3cm}
\textbf{\textit{Simulation setup}}--We approximate the background as Minkowski spacetime and adopt the simplified scalar potential
\begin{align}
      V(\phi, T) =& \frac{1}{2} \gamma (T^2-T_{0}^2) \phi^2 - \frac{1}{3}\alpha T \phi^3 + \frac{1}{4} \lambda \phi^4 \notag,
\end{align}
where we fix $\gamma = 1.11$, $\alpha = 1.41$, $\lambda = 0.5$ and $T_0 = 30 ~\mathrm{GeV}$, so that the potential has two local minimums: the symmetric vacuum $\phi_s = 0~{\rm GeV}$ and a broken-symmetry vacuum $\phi_b$. To investigate the impact of thermal fluctuations on bubble dynamics, we evolve the scalar field according to 
\begin{equation}\label{eq: bubble evolve}
    \frac{\partial ^2 \phi}{\partial t^2} - \nabla^2 \phi = -\frac{\partial V(\phi)}{\partial \phi},
\end{equation}
with initial conditions $\phi(\bold{x}, 0) = \phi_{\rm bounce} + \delta\phi$ 
where $\phi_{\rm bounce}$ is the spherically symmetric bounce profile computed using the shooting method~\cite{Athron:2024xrh} and $\delta \phi$ denotes the correction induced by thermal fluctuations. This initial condition is physically motivated by the fact that thermal bubble nucleation necessarily takes place in a thermal bath, rather than as an isolated configuration. The critical bubble itself can be regarded as a rare, large-amplitude thermal fluctuation. As a result, the post-nucleation bubble profile is generically embedded in a fluctuating background rather than being perfectly spherically symmetric. We adopt the following thermal spectrum~\cite{Bian:2025twi,Pirvu:2023plk} to
sampling thermal fluctuations
\begin{align}  \label{eq: thermal specturm}
    \bra{0}\delta\phi(\bold{k})\delta \phi^{*}(\bold{k'})\ket{0}  &= \frac{(2\pi)^3\delta^{3}(\bold{k}-\bold{k'})}{w_k} \frac{1}{e^{\frac{w_k}{T}}-1},\\
\bra{0}\delta \pi(\bold{k}) \delta \pi^{*}(\bold{k'})\ket{0} &= (2\pi)^3\delta^{3}(\bold{k}-\bold{k'})  \frac{w_k}{e^{\frac{w_k}{T}}-1}, \\
\bra{0}\delta\phi(\bold{k})\delta \pi^{*}(\bold{k'})\ket{0} &= 0,
\end{align}
with $w_k = \sqrt{k^2 + m^2}$ and $m^2= \frac{d^2 V}{d \phi^2}|_{\phi = \phi_s}$.

The simulations are performed under the following dimensionless conventions to improve numerical stability
\[
\renewcommand{\arraystretch}{1.7}
\begin{array}{lllll}
\displaystyle \bar{\phi} = \frac{\phi}{\phi_b},~
 \displaystyle \delta\bar{\phi} = \frac{\delta \phi}{\phi_b},~
 M^2= \gamma(T^2-T_0^2),~ \displaystyle \bar{V} = \frac{V(\phi)}{M^2 \phi_b^2}, \\[4pt]
  \displaystyle \bar{t} = M t,~\displaystyle \bar{\mathbf{x}} = M \mathbf{x},~ \displaystyle \bar{\pi} = \frac{\pi}{M \phi_b},~ \displaystyle \delta \bar{\pi} = \frac{\delta \pi}{M \phi_b},~
 \displaystyle \bar{\mathbf{k}} = \frac{\mathbf{k}}{M}.
\end{array}
\]
For simplicity, we assume simultaneous nucleation of the bubbles. 
We solve the real-time dynamics using a leap-frog integrator, and the GW spectrum is evaluated using the \texttt{CosmoLattice} procedure~\cite{Figueroa:2020rrl, Figueroa:2021yhd}. To enhance computational performance, the code employs 32-bit floats and is implemented in \texttt{Taichi Lang} language to exploit CUDA parallelism~\cite{hu2019taichi}. We perform lattice simulations with spatial spacing $d\bar{x} = 0.22$, and time step 
$d\bar{t}=d\bar{x}/5$, at temperatures $T=\{50~\mathrm{GeV},55~\mathrm{GeV},60~\mathrm{GeV}\}$. Single bubble dynamics is simulated on $256^3$ lattice, while GW spectra are computed using  $256^3$ and $512^3$ lattice to ensure adequate infrared resolution. Spatial discretization $d\bar{x}$ is selected to prevent artificial ultraviolet features in the GW spectrum~\cite{Dankovsky:2024zvs}. Periodic boundary conditions are adopted throughout. 

\vspace{.3cm}
\textbf{\textit{Numerical results}}--We begin by initializing a single bubble at the center of $256^3$ lattice to examine how thermal fluctuations affect its post-nucleation evolution.
\begin{figure}[htbp!]
\centering 
\includegraphics[width=0.48\textwidth]{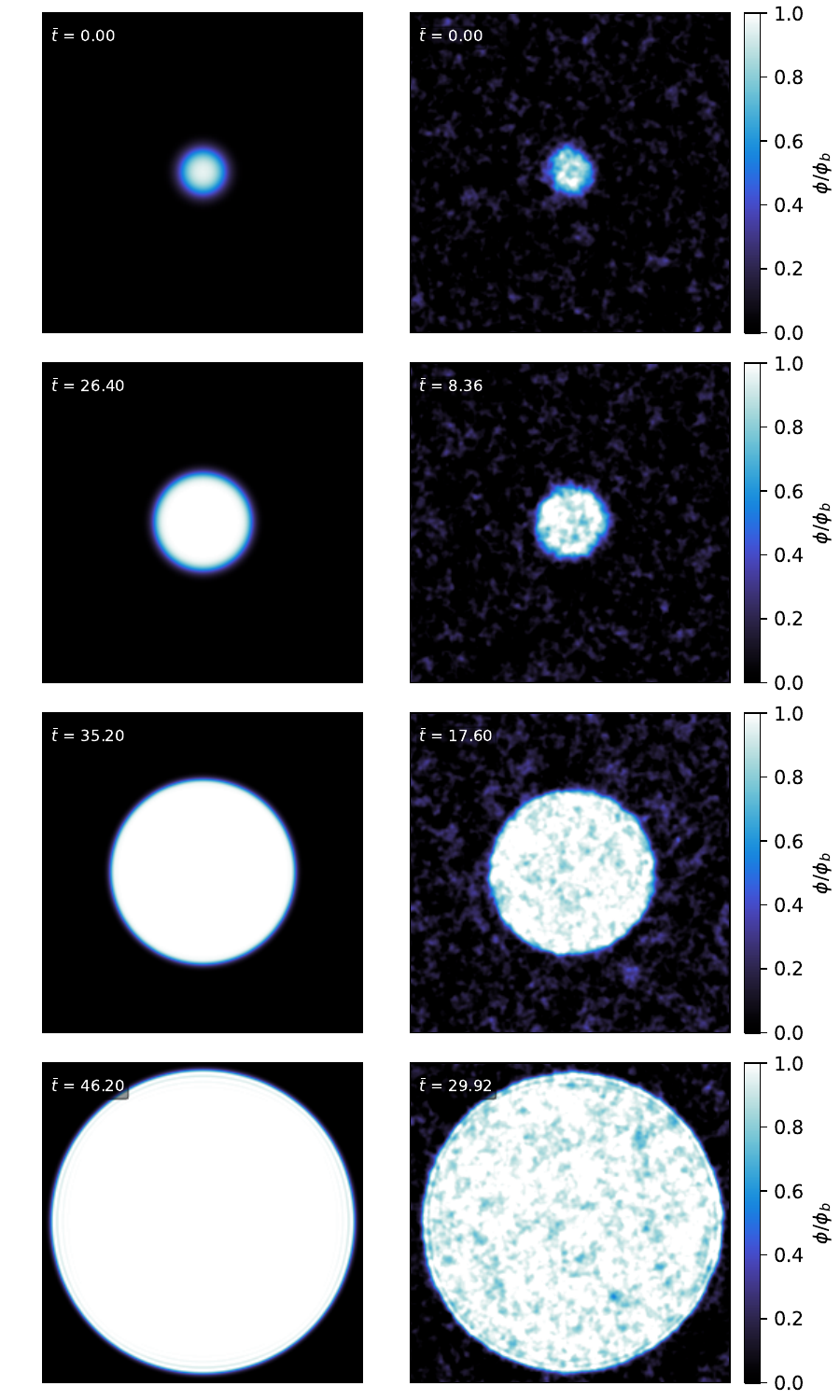}
\vspace*{-.5cm}
\caption{Slices of $\phi$ along the $z$-direction at selected times for the fluctuation-free case (left) and the case with thermal fluctuations at $T = 50~\mathrm{GeV}$ (right). The bubble is initialized on a $256^3$ lattice at $t=0$. For visual clarity, the colorbar is clipped to $[0,1]$.} 
\label{fig: fluctuation-free-bubble_evolve}
\end{figure}
In particular, we focus on the case with thermal fluctuations at $T = 50~\mathrm{GeV}$, shown in comparison with the fluctuation-free case in Fig.~\eqref{fig: fluctuation-free-bubble_evolve}. Owing to the fluctuations,  the initial bubble is distinctly irregular and exhibits a clear departure from spherical symmetry. However, this geometric distortion gradually diminishes as the bubble radius increases. This trend consistent with earlier studies based on the Nambu–Goto effective action~\cite{Adams:1989su}. 
Note that the dynamics do not completely erase the imprint of the initial fluctuations, as the interior field value remains anisotropic even at $\bar{t} \approx 30$.
Compared with the fluctuation-free case, thermal fluctuations accelerate bubble growth by truncating the initial slow-growth stage, reducing the filling time of the simulation volume by a factor of 1.6. In addition, they excite the translational mode and induce a small drift of the bubble center, visible at $\bar{t} \approx 30$.


To make these observations quantitative, we introduce a metric that tracks the loss of spherical symmetry of the bubble over time. 
Given that the relevant dynamics are governed by the wall region, it is natural to characterize the bubble geometry by defining a field-based radius $R(\hat{\mathbf n},\bar t)$ along the direction $\hat{\mathbf n}$ at time $\bar t$. This radius is specified by the condition that the field value first crosses the threshold:
\begin{equation}
  \bar{\phi}~\!\big(R(\hat{\mathbf n},\bar t)\,\hat{\mathbf n},\,\bar t\big)=\alpha,
\end{equation}
where we set the threshold $\alpha$ as 0.65. Then the symmetry breaking can be quantified by the directional variation of $R$, and we introduce a spherical symmetry factor
\begin{equation}
    \epsilon(\bar t) = \frac{{\rm Std}[R(\hat{\mathbf n},\bar t)]}{{\rm Avg}[R(\hat{\mathbf n},\bar t)]},
\end{equation}
with $ \hat{\mathbf n} \in \{ \pm \hat{\mathbf x}, \pm \hat{\mathbf y},  \pm \hat{\mathbf z}\}$,  $\mathrm{Std}$ denotes the standard deviation of these six radii at $\bar t$ and $\mathrm{Avg}$ denotes their mean. Fluctuations may perturb the bubble center, which we correct by implementing a weighted barycentric estimator. By definition, $\epsilon$ vanishes under perfect spherical symmetry and grows with increasing deviation from sphericity. Further details on this factor are provided in the \textit{Supplemental Material}. 

\begin{figure}[htbp!]
\centering 
\includegraphics[width=0.48\textwidth]{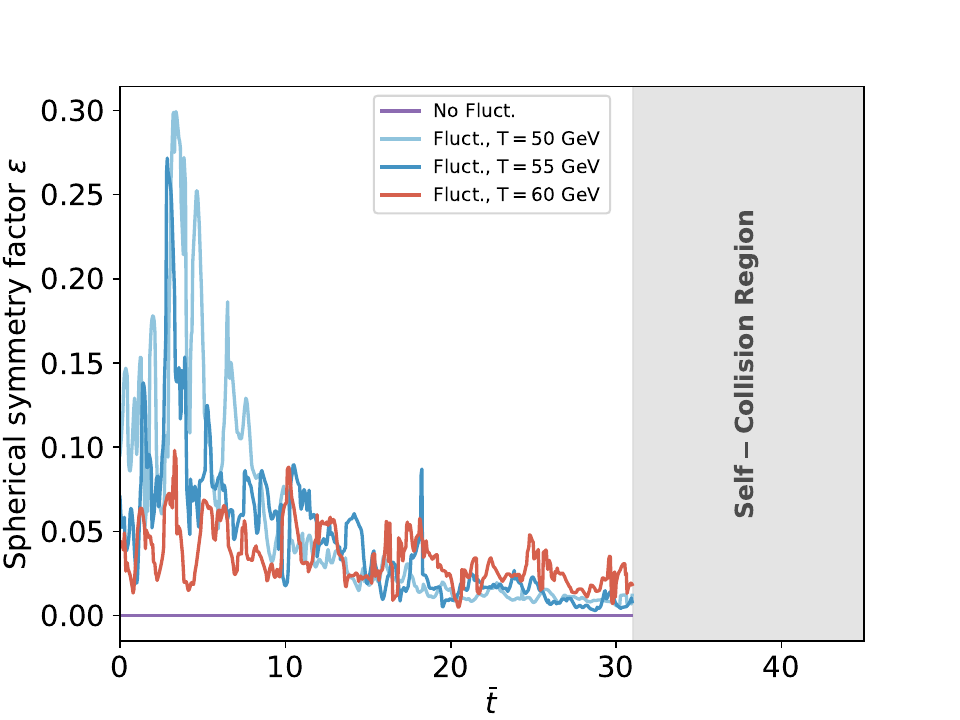}
\vspace*{-.7cm}
\caption{Time evolution of the symmetry factor $\epsilon$ computed on a $256^3$ lattice for different temperatures. 
The purple curve shows the no-fluctuation case, while the light blue, blue, and red curves correspond to fluctuations at $T = 50~\mathrm{GeV}$, $55~\mathrm{GeV}$, and $60~\mathrm{GeV}$. 
The shaded gray region marks the interval influenced by bubble self-collisions arising from periodic boundary conditions.}
\label{fig: spherical_symmetry_evolve}
\end{figure}

The time evolution of $\epsilon$ at different temperatures is shown in Fig.~\eqref{fig: spherical_symmetry_evolve}.
In the absence of fluctuations, $\epsilon\approx 0$ up to $\bar{t} \approx 50$. For $\bar{t}>50$, periodic boundaries induce self-collisions of the bubble, which break spherical symmetry and generate a nonzero $\epsilon$. Here, we focus only on the bubble expansion process, so the numerical error in $\epsilon$ is negligible. 
With fluctuations present, the value of $\epsilon$ mirrors the behavior observed in the spatial slices shown in Fig.~\eqref{fig: fluctuation-free-bubble_evolve}.
We find that the initial fluctuation-induced distortion becomes more pronounced at lower temperatures. 
This counterintuitive trend can be attributed to the fact that, within our simplified potential, the barrier decreases as $T$ drops.
Thus, even though thermal fluctuations are weaker at lower $T$, the bubble becomes more susceptible to perturbations.

The emergence of bubble asphericity generates a nonzero quadrupole moment and thus induces gravitational radiation. We compute the normalized GW spectrum for a single bubble in the presence of fluctuations at $T = 50 ~\mathrm{GeV}$~\cite{Cutting:2018tjt,Cutting:2020nla} 
\begin{align}
    \frac{1}{\left(HR_{*}\Omega_{\rm vac}\right)^2}\frac{d \Omega_{\rm GW}}{d lnk} = \frac{1}{\left(HR_{*}\Omega_{\rm vac}\right)^2 \rho_c}\frac{d \rho_{\rm GW}}{d lnk},
\end{align}
where $\rho_{\rm gw}$ is the energy density of GW, $\rho_c \equiv 3H^2/(8\pi G)$ is the critical density, $R_{*}$ denotes the characteristic length scale of the FOPT, and $\Omega_{\rm vac}$ is the potential energy difference between the two vacua normalized by the critical density.

\begin{figure}[!htbp]
\centering 
\includegraphics[width=0.48\textwidth]{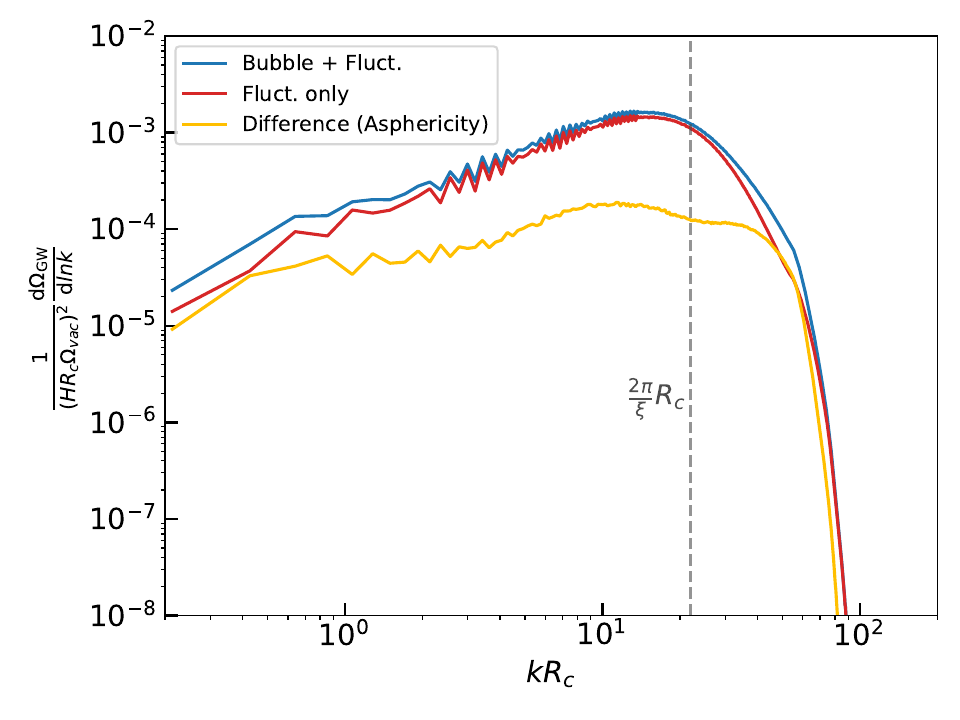}
\vspace*{-.9cm}
\caption{GW spectra computed at $\bar t = 30$ 
for a single bubble with thermal fluctuations at $T = 50~\mathrm{GeV}$.
The blue curve shows the result from evolving the full scalar field, the red curve corresponds to the evolution of fluctuations alone, and the orange curve denotes their difference, which can be interpreted as the contribution from the evolution of a non-spherical bubble.}
\label{fig: single_bubble_GW}
\end{figure}

The results obtained on a $512^3$ lattice are shown in Fig.~\eqref{fig: single_bubble_GW}.
We also calculate the fluctuation-free case to establish a numerical error baseline. For single bubble simulation, we choose the bubble radius $R_c$ as the characteristic length scale, defined by $\bar{\phi}(R_c, \bar{t}=0)= \bar{\phi}_b/2$. At $\bar t=30$, the resulting GW spectrum peaks at the $\sim10^{-9}$ level which lies far below the lower limit of the plot. This is consistent with numerical noise given that an isolated spherical bubble does not emit GW prior to collisions. By contrast, once fluctuations are included, it shows that the GW spectrum from the full scalar-field
evolution peaks at the $\sim 10^{-3}$ level. In principle, thermal fluctuations outside the bubble region also source GWs. We therefore subtract the spectrum obtained by evolving the same fluctuations without a bubble, so that the residual signal isolates the bubble–fluctuation coupling, i.e., the GW emitted by a non-spherical bubble. Evaluated at $\bar{t}=30$, this residual spectrum features a prominent peak at $kR_c\simeq 10$. Since thermal fluctuations have a characteristic length $\xi\simeq 1/T$, the corresponding dimensionless wavenumber can be estimated as
\begin{equation}
kR_c \sim \frac{2\pi}{\xi}R_c \simeq 2\pi T R_c \approx 21.9,
\end{equation}
which matches the location of the high-$k$ feature and indicates that it is primarily sourced by small-scale structures seeded by thermal fluctuations.

\begin{figure}[htbp!]
\centering 
\includegraphics[width=0.5\textwidth]{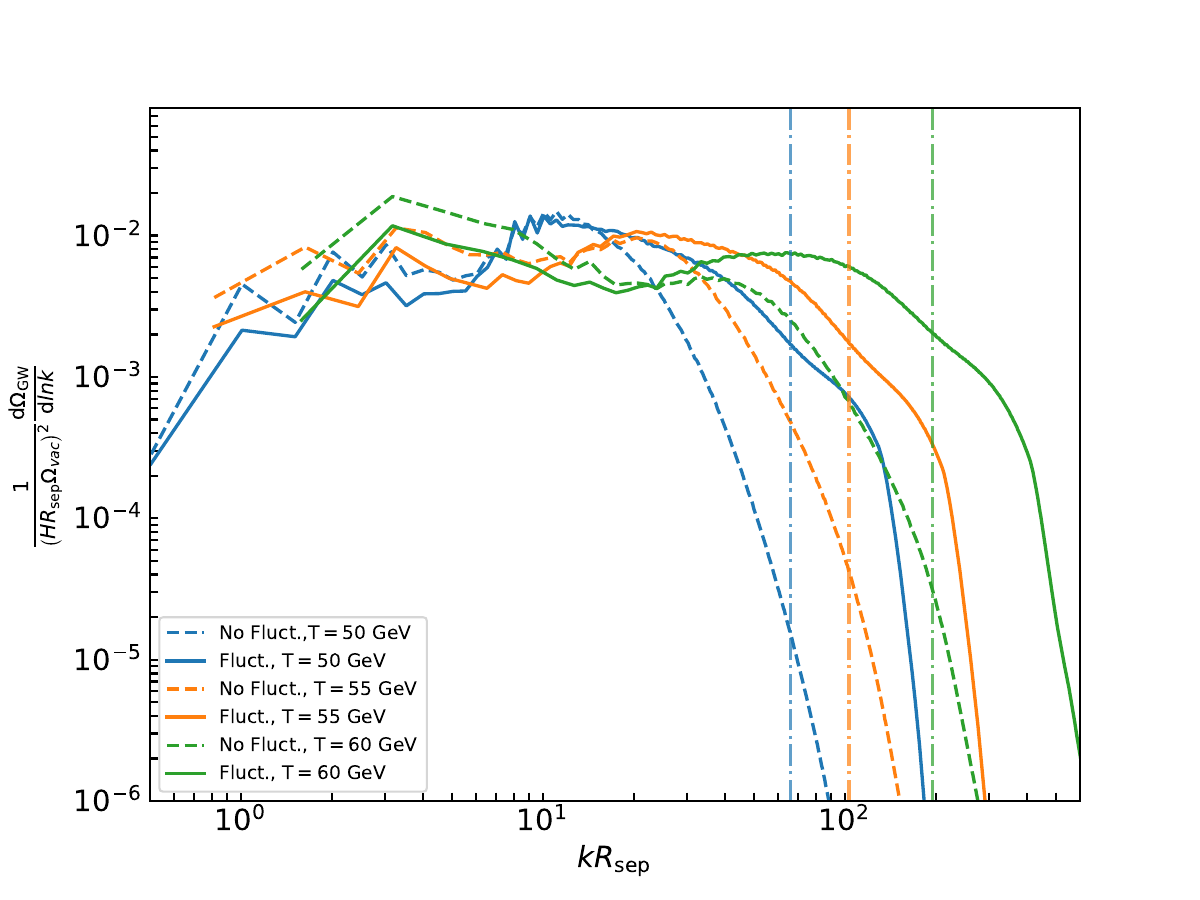}
\vspace*{-.8cm}
\caption{GW spectra computed on a $512^3$ lattice for multi-bubble evolution, evaluated at $\bar{t} = 4 R_{\mathrm{sep}} M$, for $T = 50~\mathrm{GeV}$, $55~\mathrm{GeV}$, and $60~\mathrm{GeV}$ (blue, green, and orange). Dashed lines denote the corresponding spectra without thermal fluctuations. The vertical dash–dot lines mark the characteristic scales $kR_{\mathrm{sep}} = 2\pi R_{\mathrm{sep}}T$.}

\label{fig: multi_bubble_GW}
\end{figure}

We now examine how these effects manifest in the GW spectrum from multiple bubbles.
For our simulation parameters, a volume of $(512 d\bar{x})^3$ can accommodate up to 60--2000 bubbles, depending on the bubble size. Therefore, we choose the characteristic length scale to be the mean bubble separation, denoted by $R_{\mathrm{sep}}$. The final results at $\bar{t} = 4R_{\mathrm{sep}}$ are shown in Fig.~\eqref{fig: multi_bubble_GW}. 

Compared to the case without fluctuation, thermal fluctuations modify the infrared behavior and reduce the spectral amplitude around $kR_{\mathrm{sep}}\approx 2\pi$. By contrast, the ultraviolet part of the spectrum is significantly enhanced and falls off more slowly than the one without fluctuations. A simple dimensional estimate of the fluctuation contribution places this enhancement at wavenumbers, $kR_{\mathrm{sep}} \approx \frac{2 \pi}{\xi} R_{\mathrm{sep}}$, consistent with the observed shift of the ultraviolet features. With higher temperatures, the ultraviolet enhancement becomes larger, suggesting that comparison with the fluctuation-free baseline may allow one to infer the approximate transition temperature and thereby extract additional information from the GW signal. 
Moreover, the GW spectrum in the presence of thermal fluctuations develops a clear break in the spectral slope in the ultraviolet regime, around a characteristic dimensionless wavenumber $k_{ b}R_{\rm sep}$. In our simulations, we find
$k_bR_{\rm sep} \approx 100, 200, 400$  
for $T=50~{\rm GeV}$, $55~{\rm GeV}$, and $60~{\rm GeV}$, respectively.

To providing a qualitative characterization of the spectral behavior, we analyze it from the perspective of the underlying source. Given the GW spectrum is governed by the correlator of the transverse--traceless (TT) anisotropic stress, we focus on the parametric $k$-dependence of
$\langle \Pi^{\rm tot}_{ij}(t,\mathbf{k})\,\Pi^{\rm tot}_{ij}(t,\mathbf{k})\rangle$. In this analysis we retain only the leading exponential envelopes in the ultraviolet and treating the TT projection and convolution structure as contributing at most algebraic prefactors in $k$. In our simulations, the stress can be decomposed as
\begin{equation}
\Pi^{\rm tot}_{ij}=\Pi^{bb}_{ij}+\Pi^{b\delta}_{ij}+\Pi^{\delta\delta}_{ij},
\end{equation}
corresponding to $(\nabla\phi_b)^2$, $(\nabla\phi_b)(\nabla\delta\phi)$, and $(\nabla\delta\phi)^2$, respectively. Approximating the bubble profile by a $\tanh$ wall with effective width $l_w$, its Fourier transform suggests the envelope estimate
\begin{equation} \label{eq: bb correlation}
\left\langle \Pi^{bb}_{ij}(t,\mathbf{k})\,\Pi^{bb}_{ij}(t,\mathbf{k})\right\rangle
\ \propto\ e^{-l_w k}\,
\end{equation}
which accounts for the exponential high-$k$ tail indicated by dashed line in Fig.~\eqref{fig: multi_bubble_GW}, but not include the collision contribution.

Incorporating the thermal fluctuations come in and following Eq.~\eqref{eq: thermal specturm}, we estimate
\begin{equation}
    \begin{aligned}
        \left\langle \Pi^{b\delta}_{ij}(t,\mathbf{k})\,\Pi^{b\delta}_{ij}(t,\mathbf{k})\right\rangle 
        &\propto \left<\phi_b \phi_b\right> \left<\delta \phi \delta \phi\right> \\
        &\propto  e^{-(l_w + \delta l_w)k}e^{-k/T},  \\
        \left\langle \Pi^{\delta\delta}_{ij}(t,\mathbf{k})\,\Pi^{\delta\delta}_{ij}(t,\mathbf{k})\right\rangle 
        &\propto  \left<\delta \phi \delta \phi\right> \left<\delta \phi \delta \phi\right>  \\ 
        &\propto  e^{-2k/T},
    \end{aligned}
\end{equation}
and other six unlisted cross-terms in the Boltzmann regime $k\gtrsim T$. These expressions highlights that the ultraviolet behavior is controlled by the interplay of three characteristic scales: the bubble-wall thickness $l_w$, the deformation scale of the bubble profile $\delta l_w$, and
the fluctuation length $1/T$. 
In the high-$k$ regime, the dominant contribution is expected to originate from the smallest length scales, i.e., from pure fluctuations. We therefore anticipate
\begin{equation}\label{eq: guess equation}
    \begin{aligned}
        \mathcal R - 1 &\ \ \equiv\ \left(\frac{\mathrm d\Omega^{\rm fluc.}_{\rm GW}}{\mathrm d\ln k}\right)\bigg/\left(\frac{\mathrm d\Omega^{\rm No\ fluc.}_{\rm GW}}{\mathrm d\ln k}\right) - 1 \\
        &\overset{\text{high-$k$}}{\propto} \exp\!\left[\left(\frac{l_w}{R_{\rm sep}}-\frac{2}{T R_{\rm sep}}\right)kR_{\rm sep}\right].
    \end{aligned}
\end{equation}

\begin{table}[]
\begin{tabular}{cccccc}
\hline
$T(\mathrm{GeV})$  & $R_{\rm sep}(\mathrm{GeV})^{-1}$    & $2/(T R_{\rm sep})$      & $l_w/R_{\rm sep}$ & slope & Relative Error \\
\hline
50 & 0.210 & 0.190 & 0.128 & -0.088 & 30\%  \\
55 & 0.301 & 0.121  & 0.065 & -0.059 & 8\%   \\
60 & 0.520 & 0.064   & 0.036 & -0.024 & 17\% \\
\hline
\end{tabular}
\caption{Key quantities extracted from the simulations and corresponding fit results.}
\label{tab: simulation data}
\end{table}
The slope of $\ln(\mathcal{R}-1)$ in the post-break region, $kR_{\rm sep}>k_bR_{\rm sep}$, together with the ratio $l_w/R_{\rm sep}$, can be extracted directly from our GW simulations. The results are summarized in Tab.~\ref{tab: simulation data}. These values exhibit good order-of-magnitude agreement with Eq.~\eqref{eq: guess equation}, with relative deviations of $30\%$, $8\%$, and $17\%$ for $T=50$, $55$, and $60~{\rm GeV}$, respectively. The larger discrepancy at $T=50~{\rm GeV}$ is plausibly attributable to the thinner bubble wall and the reduced scale separation, such that the high-$k$ tail is not yet purely fluctuation-dominated.

\begin{figure}[htbp!]
\centering 
\includegraphics[width=0.5\textwidth]{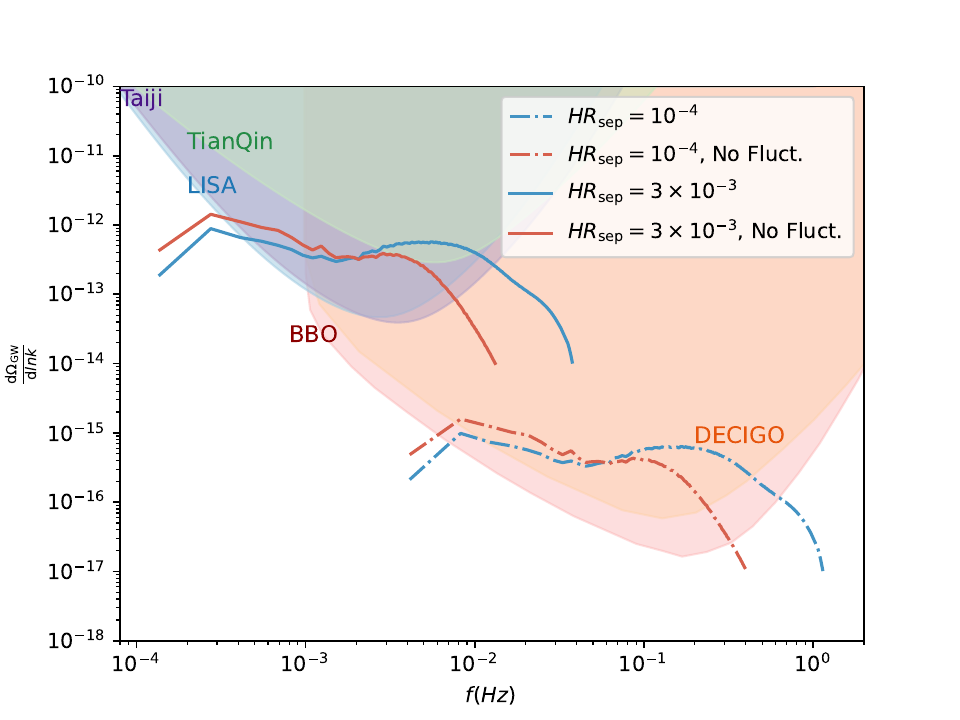}
\vspace*{-.7cm}
\caption{Detectability of the GW spectra extrapolated from simulations performed at $T= 60~\mathrm{GeV}$ and $\bar{t}=4R_{\mathrm{sep}}M$.
The solid curves correspond to $HR_{\mathrm{sep}}=0.003$, while the dashed curves denote $HR_{\mathrm{sep}}=0.0001$.
The sensitivity curves of LISA, Taiji, TianQin, DECIGO~\cite{Schmitz:2020syl}, and BBO~\cite{Schmitz:2020syl} are shown for comparison. The sensitivity curves for GW experiments are for 1 year observation time, with signal-to-noise ratio = 1.} 
\label{fig: GW_exp}
\end{figure}

To assess the detectability of the thermal fluctuation induced modifications shown in Fig.~\eqref{fig: multi_bubble_GW}, we extrapolate our normalized GW spectra, evaluated at $\bar{t} = 4R_{\mathrm{sep}}M$ and at a temperature $T = 60~\mathrm{GeV}$, to physical cosmological scales. In our normalization, the overall scale dependence is factored out, so that recovering the physical frequency range and the amplitude of the signal requires specifying the dimensionless parameters $HR_{\mathrm{sep}}$ and $\Omega_{\rm vac}$. Then the physical frequency is given by
\begin{equation}
    f(k) = 2.6 \times 10^{-6} \mathrm{Hz} \frac{kR_{\mathrm{sep}}}{HR_{\mathrm{sep}}}\frac{T}{100~\mathrm{GeV}} \left(\frac{g_{*}}{100}\right)^{1/6},
\end{equation}
and the amplitude of the signal is scaled by $(HR_{\mathrm{sep}}\Omega_{\mathrm{vac}})^2$. For $\Omega_{\rm vac}$, we adopt as $0.2\%$ predicted by our toy model. The parameter $HR_{\mathrm{sep}}$ is related to the inverse duration of the phase transition via $\beta R_{\mathrm{sep}} = (8\pi)^{1/3} v_w$~\cite{Hindmarsh:2019phv}, and for electroweak FOPT one typically has $\beta/H \sim 10^{2}\text{--}10^{4}$. In this parameter range, we find that over a broad range of $HR_{\mathrm{sep}}$, the corresponding GW signals are detectable by future experiments. In Fig.~\ref{fig: GW_exp}, we present the cases with $HR_{\mathrm{sep}}=3 \times 10^{-3}$ and $10^{-4}$. For the smaller value $HR_{\mathrm{sep}}=10^{-4}$, the impact of thermal fluctuations is potentially detectable by BBO~\cite{Harry_2006,Corbin_2006,PhysRevD.72.083005} and DECIGO~\cite{Kawamura_2011,10.1093/ptep/pty078, PhysRevD.83.044011, PhysRevLett.87.221103}, whereas for $HR_{\mathrm{sep}}=3 \times 10^{-3}$, the same effects fall within the sensitivity ranges of LISA~\cite{amaro2017laser,Baker:2019nia}, Taiji~\cite{taiji,taijii}, and TianQin~\cite{Luo_2016, Zhou:2023rop}. 

\vspace{.2cm}
\textbf{\textit{Conclusion}}-- In summary, we performed $(3+1)$-dimensional lattice simulations of a scalar-field model with thermal initial conditions to quantify how thermal fluctuations initially distort spherical vacuum bubbles and reshape the resulting GW spectrum. We quantitatively showed that thermal fluctuations can induce substantial early-time asphericity of the bubble wall, such that even an isolated bubble can become a transient GW source. In multi-bubble simulations, thermal fluctuations primarily redistribute power across scales. they suppress the low-$k$ region while enhancing the high-$k$ tail in a temperature-dependent manner, yielding spectra that differ qualitatively from fluctuation-free expectations. Extrapolating to cosmological scales, we found that for a broad range of $H R_{\rm sep}$ the resulting signals can lie within the reach of future GW observatories, and the fluctuation-induced reshaping of the spectral morphology may be important for the identification and interpretation of GW from electroweak-scale phase transitions.

These results also motivate an extension to a more realistic thermal plasma environment. In cosmological FOPTs, a dominant contribution to the GW background is sourced by long-lived acoustic motions in the surrounding fluid. If thermal fluctuations break the spherical symmetry of bubble walls, they should naturally induce non-spherically symmetric fluid profiles in the near-wall region, thereby modifying the acoustic source and potentially imprinting fluctuation-induced signatures on the sound-wave GW spectrum. Thermal fluctuation induced asphericity may also affect electroweak baryogenesis by making local wall velocity anisotropic, $v_w \to v_w(r,\theta,\varphi)$, thereby rendering the CP-violating sources and transport dynamics inhomogeneous. This introduces additional theoretical uncertainty beyond the standard assumption of a uniform wall velocity.
         
\vspace{.2cm}
\textit{Acknowledgments}--We thank Bing Sun for providing access to high-performance computing resources in the early stages of this project. We also thank Zheng-Cheng Liang for providing the sensitivity curves for LISA and TianQin, and Huai-ke Guo for providing the sensitivity curve for Taiji. The numerical calculations in this work were carried out on the High-Performance Computing Platform at the Center for Theoretical Physics, Henan Normal University. This work was supported by the National Natural Science Foundation of China (Grant Nos. 12335005, 12322505, and 12547101), the PI Research Fund of Henan Normal University (Grant No. 5101029470335), the Chongqing Natural Science Foundation (Grant No. CSTB2024NSCQ-JQX0022), and the Chongqing Talents: Exceptional Young Talents Project (No. cstc2024ycjh-bgzxm0020).


\bibliography{ref}

\begin{thebibliography}{45}
\expandafter\ifx\csname natexlab\endcsname\relax\def\natexlab#1{#1}\fi
\expandafter\ifx\csname bibnamefont\endcsname\relax
  \def\bibnamefont#1{#1}\fi
\expandafter\ifx\csname bibfnamefont\endcsname\relax
  \def\bibfnamefont#1{#1}\fi
\expandafter\ifx\csname citenamefont\endcsname\relax
  \def\citenamefont#1{#1}\fi
\expandafter\ifx\csname url\endcsname\relax
  \def\url#1{\texttt{#1}}\fi
\expandafter\ifx\csname urlprefix\endcsname\relax\def\urlprefix{URL }\fi
\providecommand{\bibinfo}[2]{#2}
\providecommand{\eprint}[2][]{\url{#2}}

\bibitem[{\citenamefont{Morrissey and Ramsey-Musolf}(2012)}]{Morrissey:2012db}
\bibinfo{author}{\bibfnamefont{D.~E.} \bibnamefont{Morrissey}} \bibnamefont{and} \bibinfo{author}{\bibfnamefont{M.~J.} \bibnamefont{Ramsey-Musolf}}, \bibinfo{journal}{New J. Phys.} \textbf{\bibinfo{volume}{14}}, \bibinfo{pages}{125003} (\bibinfo{year}{2012}), \eprint{1206.2942}.

\bibitem[{\citenamefont{Kuzmin et~al.}(1985)\citenamefont{Kuzmin, Rubakov, and Shaposhnikov}}]{Kuzmin:1985mm}
\bibinfo{author}{\bibfnamefont{V.~A.} \bibnamefont{Kuzmin}}, \bibinfo{author}{\bibfnamefont{V.~A.} \bibnamefont{Rubakov}}, \bibnamefont{and} \bibinfo{author}{\bibfnamefont{M.~E.} \bibnamefont{Shaposhnikov}}, \bibinfo{journal}{Phys. Lett. B} \textbf{\bibinfo{volume}{155}}, \bibinfo{pages}{36} (\bibinfo{year}{1985}).

\bibitem[{\citenamefont{Kawamura et~al.}(2011)\citenamefont{Kawamura, Ando, Seto, Sato, Nakamura, Tsubono, Kanda, Tanaka, Yokoyama, Funaki et~al.}}]{Kawamura_2011}
\bibinfo{author}{\bibfnamefont{S.}~\bibnamefont{Kawamura}}, \bibinfo{author}{\bibfnamefont{M.}~\bibnamefont{Ando}}, \bibinfo{author}{\bibfnamefont{N.}~\bibnamefont{Seto}}, \bibinfo{author}{\bibfnamefont{S.}~\bibnamefont{Sato}}, \bibinfo{author}{\bibfnamefont{T.}~\bibnamefont{Nakamura}}, \bibinfo{author}{\bibfnamefont{K.}~\bibnamefont{Tsubono}}, \bibinfo{author}{\bibfnamefont{N.}~\bibnamefont{Kanda}}, \bibinfo{author}{\bibfnamefont{T.}~\bibnamefont{Tanaka}}, \bibinfo{author}{\bibfnamefont{J.}~\bibnamefont{Yokoyama}}, \bibinfo{author}{\bibfnamefont{I.}~\bibnamefont{Funaki}}, \bibnamefont{et~al.}, \bibinfo{journal}{Classical and Quantum Gravity} \textbf{\bibinfo{volume}{28}}, \bibinfo{pages}{094011} (\bibinfo{year}{2011}), \urlprefix\url{https://doi.org/10.1088/0264-9381/28/9/094011}.

\bibitem[{\citenamefont{Baker and Mittnacht}(2019)}]{Baker:2018vos}
\bibinfo{author}{\bibfnamefont{M.~J.} \bibnamefont{Baker}} \bibnamefont{and} \bibinfo{author}{\bibfnamefont{L.}~\bibnamefont{Mittnacht}}, \bibinfo{journal}{JHEP} \textbf{\bibinfo{volume}{05}}, \bibinfo{pages}{070} (\bibinfo{year}{2019}), \eprint{1811.03101}.

\bibitem[{\citenamefont{Baker et~al.}(2020)\citenamefont{Baker, Kopp, and Long}}]{Baker:2019ndr}
\bibinfo{author}{\bibfnamefont{M.~J.} \bibnamefont{Baker}}, \bibinfo{author}{\bibfnamefont{J.}~\bibnamefont{Kopp}}, \bibnamefont{and} \bibinfo{author}{\bibfnamefont{A.~J.} \bibnamefont{Long}}, \bibinfo{journal}{Phys. Rev. Lett.} \textbf{\bibinfo{volume}{125}}, \bibinfo{pages}{151102} (\bibinfo{year}{2020}), \eprint{1912.02830}.

\bibitem[{\citenamefont{Baker et~al.}(2018)\citenamefont{Baker, Breitbach, Kopp, and Mittnacht}}]{Baker:2017zwx}
\bibinfo{author}{\bibfnamefont{M.~J.} \bibnamefont{Baker}}, \bibinfo{author}{\bibfnamefont{M.}~\bibnamefont{Breitbach}}, \bibinfo{author}{\bibfnamefont{J.}~\bibnamefont{Kopp}}, \bibnamefont{and} \bibinfo{author}{\bibfnamefont{L.}~\bibnamefont{Mittnacht}}, \bibinfo{journal}{JHEP} \textbf{\bibinfo{volume}{03}}, \bibinfo{pages}{114} (\bibinfo{year}{2018}), \eprint{1712.03962}.

\bibitem[{\citenamefont{Hawking et~al.}(1982)\citenamefont{Hawking, Moss, and Stewart}}]{Hawking:1982ga}
\bibinfo{author}{\bibfnamefont{S.~W.} \bibnamefont{Hawking}}, \bibinfo{author}{\bibfnamefont{I.~G.} \bibnamefont{Moss}}, \bibnamefont{and} \bibinfo{author}{\bibfnamefont{J.~M.} \bibnamefont{Stewart}}, \bibinfo{journal}{Phys. Rev. D} \textbf{\bibinfo{volume}{26}}, \bibinfo{pages}{2681} (\bibinfo{year}{1982}).

\bibitem[{\citenamefont{Carr et~al.}(2021)\citenamefont{Carr, Kohri, Sendouda, and Yokoyama}}]{Carr:2020gox}
\bibinfo{author}{\bibfnamefont{B.}~\bibnamefont{Carr}}, \bibinfo{author}{\bibfnamefont{K.}~\bibnamefont{Kohri}}, \bibinfo{author}{\bibfnamefont{Y.}~\bibnamefont{Sendouda}}, \bibnamefont{and} \bibinfo{author}{\bibfnamefont{J.}~\bibnamefont{Yokoyama}}, \bibinfo{journal}{Rept. Prog. Phys.} \textbf{\bibinfo{volume}{84}}, \bibinfo{pages}{116902} (\bibinfo{year}{2021}), \eprint{2002.12778}.

\bibitem[{\citenamefont{Athron et~al.}(2024)\citenamefont{Athron, Bal{\'a}zs, Fowlie, Morris, and Wu}}]{Athron:2023xlk}
\bibinfo{author}{\bibfnamefont{P.}~\bibnamefont{Athron}}, \bibinfo{author}{\bibfnamefont{C.}~\bibnamefont{Bal{\'a}zs}}, \bibinfo{author}{\bibfnamefont{A.}~\bibnamefont{Fowlie}}, \bibinfo{author}{\bibfnamefont{L.}~\bibnamefont{Morris}}, \bibnamefont{and} \bibinfo{author}{\bibfnamefont{L.}~\bibnamefont{Wu}}, \bibinfo{journal}{Prog. Part. Nucl. Phys.} \textbf{\bibinfo{volume}{135}}, \bibinfo{pages}{104094} (\bibinfo{year}{2024}), \eprint{2305.02357}.

\bibitem[{\citenamefont{Coleman}(1977)}]{Coleman:1977py}
\bibinfo{author}{\bibfnamefont{S.~R.} \bibnamefont{Coleman}}, \bibinfo{journal}{Phys. Rev. D} \textbf{\bibinfo{volume}{15}}, \bibinfo{pages}{2929} (\bibinfo{year}{1977}), \bibinfo{note}{[Erratum: Phys.Rev.D 16, 1248 (1977)]}.

\bibitem[{\citenamefont{Coleman et~al.}(1978)\citenamefont{Coleman, Glaser, and Martin}}]{Coleman:1977th}
\bibinfo{author}{\bibfnamefont{S.~R.} \bibnamefont{Coleman}}, \bibinfo{author}{\bibfnamefont{V.}~\bibnamefont{Glaser}}, \bibnamefont{and} \bibinfo{author}{\bibfnamefont{A.}~\bibnamefont{Martin}}, \bibinfo{journal}{Commun. Math. Phys.} \textbf{\bibinfo{volume}{58}}, \bibinfo{pages}{211} (\bibinfo{year}{1978}).

\bibitem[{\citenamefont{Linde}(1983)}]{Linde:1981zj}
\bibinfo{author}{\bibfnamefont{A.~D.} \bibnamefont{Linde}}, \bibinfo{journal}{Nucl. Phys. B} \textbf{\bibinfo{volume}{216}}, \bibinfo{pages}{421} (\bibinfo{year}{1983}), \bibinfo{note}{[Erratum: Nucl.Phys.B 223, 544 (1983)]}.

\bibitem[{\citenamefont{Hertzberg et~al.}(2020)\citenamefont{Hertzberg, Rompineve, and Shah}}]{Hertzberg:2020tqa}
\bibinfo{author}{\bibfnamefont{M.~P.} \bibnamefont{Hertzberg}}, \bibinfo{author}{\bibfnamefont{F.}~\bibnamefont{Rompineve}}, \bibnamefont{and} \bibinfo{author}{\bibfnamefont{N.}~\bibnamefont{Shah}}, \bibinfo{journal}{Phys. Rev. D} \textbf{\bibinfo{volume}{102}}, \bibinfo{pages}{076003} (\bibinfo{year}{2020}), \eprint{2009.00017}.

\bibitem[{\citenamefont{Bian et~al.}(2025)\citenamefont{Bian, Di, Jia, Li, and Zeng}}]{Bian:2025twi}
\bibinfo{author}{\bibfnamefont{L.}~\bibnamefont{Bian}}, \bibinfo{author}{\bibfnamefont{Y.}~\bibnamefont{Di}}, \bibinfo{author}{\bibfnamefont{Y.}~\bibnamefont{Jia}}, \bibinfo{author}{\bibfnamefont{Y.}~\bibnamefont{Li}}, \bibnamefont{and} \bibinfo{author}{\bibfnamefont{K.}~\bibnamefont{Zeng}} (\bibinfo{year}{2025}), \eprint{2505.15360}.

\bibitem[{\citenamefont{Batini et~al.}(2024)\citenamefont{Batini, Chatrchyan, and Berges}}]{Batini:2023zpi}
\bibinfo{author}{\bibfnamefont{L.}~\bibnamefont{Batini}}, \bibinfo{author}{\bibfnamefont{A.}~\bibnamefont{Chatrchyan}}, \bibnamefont{and} \bibinfo{author}{\bibfnamefont{J.}~\bibnamefont{Berges}}, \bibinfo{journal}{Phys. Rev. D} \textbf{\bibinfo{volume}{109}}, \bibinfo{pages}{023502} (\bibinfo{year}{2024}), \eprint{2310.04206}.

\bibitem[{\citenamefont{Braden et~al.}(2019)\citenamefont{Braden, Johnson, Peiris, Pontzen, and Weinfurtner}}]{Braden:2018tky}
\bibinfo{author}{\bibfnamefont{J.}~\bibnamefont{Braden}}, \bibinfo{author}{\bibfnamefont{M.~C.} \bibnamefont{Johnson}}, \bibinfo{author}{\bibfnamefont{H.~V.} \bibnamefont{Peiris}}, \bibinfo{author}{\bibfnamefont{A.}~\bibnamefont{Pontzen}}, \bibnamefont{and} \bibinfo{author}{\bibfnamefont{S.}~\bibnamefont{Weinfurtner}}, \bibinfo{journal}{Phys. Rev. Lett.} \textbf{\bibinfo{volume}{123}}, \bibinfo{pages}{031601} (\bibinfo{year}{2019}), \bibinfo{note}{[Erratum: Phys.Rev.Lett. 129, 059901 (2022)]}, \eprint{1806.06069}.

\bibitem[{\citenamefont{Chatrchyan et~al.}(2025)\citenamefont{Chatrchyan, Niedermann, and Richman-Taylor}}]{Chatrchyan:2025uar}
\bibinfo{author}{\bibfnamefont{A.}~\bibnamefont{Chatrchyan}}, \bibinfo{author}{\bibfnamefont{F.}~\bibnamefont{Niedermann}}, \bibnamefont{and} \bibinfo{author}{\bibfnamefont{P.}~\bibnamefont{Richman-Taylor}} (\bibinfo{year}{2025}), \eprint{2510.27579}.

\bibitem[{\citenamefont{Aguirre and Johnson}(2005)}]{Aguirre:2005xs}
\bibinfo{author}{\bibfnamefont{A.}~\bibnamefont{Aguirre}} \bibnamefont{and} \bibinfo{author}{\bibfnamefont{M.~C.} \bibnamefont{Johnson}}, \bibinfo{journal}{Phys. Rev. D} \textbf{\bibinfo{volume}{72}}, \bibinfo{pages}{103525} (\bibinfo{year}{2005}), \eprint{gr-qc/0508093}.

\bibitem[{\citenamefont{Blum and Mirbabayi}(2024)}]{Blum:2024singlebubble}
\bibinfo{author}{\bibfnamefont{K.}~\bibnamefont{Blum}} \bibnamefont{and} \bibinfo{author}{\bibfnamefont{M.}~\bibnamefont{Mirbabayi}} (\bibinfo{year}{2024}), \eprint{2403.20164}.

\bibitem[{\citenamefont{Adams et~al.}(1990)\citenamefont{Adams, Freese, and Widrow}}]{Adams:1989su}
\bibinfo{author}{\bibfnamefont{F.~C.} \bibnamefont{Adams}}, \bibinfo{author}{\bibfnamefont{K.}~\bibnamefont{Freese}}, \bibnamefont{and} \bibinfo{author}{\bibfnamefont{L.~M.} \bibnamefont{Widrow}}, \bibinfo{journal}{Phys. Rev. D} \textbf{\bibinfo{volume}{41}}, \bibinfo{pages}{347} (\bibinfo{year}{1990}).

\bibitem[{\citenamefont{Braden et~al.}(2015{\natexlab{a}})\citenamefont{Braden, Bond, and Mersini-Houghton}}]{Braden:2014cra}
\bibinfo{author}{\bibfnamefont{J.}~\bibnamefont{Braden}}, \bibinfo{author}{\bibfnamefont{J.~R.} \bibnamefont{Bond}}, \bibnamefont{and} \bibinfo{author}{\bibfnamefont{L.}~\bibnamefont{Mersini-Houghton}}, \bibinfo{journal}{JCAP} \textbf{\bibinfo{volume}{03}}, \bibinfo{pages}{007} (\bibinfo{year}{2015}{\natexlab{a}}), \eprint{1412.5591}.

\bibitem[{\citenamefont{Braden et~al.}(2015{\natexlab{b}})\citenamefont{Braden, Bond, and Mersini-Houghton}}]{Braden:2015vza}
\bibinfo{author}{\bibfnamefont{J.}~\bibnamefont{Braden}}, \bibinfo{author}{\bibfnamefont{J.~R.} \bibnamefont{Bond}}, \bibnamefont{and} \bibinfo{author}{\bibfnamefont{L.}~\bibnamefont{Mersini-Houghton}}, \bibinfo{journal}{JCAP} \textbf{\bibinfo{volume}{08}}, \bibinfo{pages}{048} (\bibinfo{year}{2015}{\natexlab{b}}), \eprint{1505.01857}.

\bibitem[{\citenamefont{Bond et~al.}(2015)\citenamefont{Bond, Braden, and Mersini-Houghton}}]{Bond:2015zfa}
\bibinfo{author}{\bibfnamefont{J.~R.} \bibnamefont{Bond}}, \bibinfo{author}{\bibfnamefont{J.}~\bibnamefont{Braden}}, \bibnamefont{and} \bibinfo{author}{\bibfnamefont{L.}~\bibnamefont{Mersini-Houghton}}, \bibinfo{journal}{JCAP} \textbf{\bibinfo{volume}{09}}, \bibinfo{pages}{004} (\bibinfo{year}{2015}), \eprint{1505.02162}.

\bibitem[{\citenamefont{Athron et~al.}(2025)\citenamefont{Athron, Balazs, Fowlie, Morris, Searle, Xiao, and Zhang}}]{Athron:2024xrh}
\bibinfo{author}{\bibfnamefont{P.}~\bibnamefont{Athron}}, \bibinfo{author}{\bibfnamefont{C.}~\bibnamefont{Balazs}}, \bibinfo{author}{\bibfnamefont{A.}~\bibnamefont{Fowlie}}, \bibinfo{author}{\bibfnamefont{L.}~\bibnamefont{Morris}}, \bibinfo{author}{\bibfnamefont{W.}~\bibnamefont{Searle}}, \bibinfo{author}{\bibfnamefont{Y.}~\bibnamefont{Xiao}}, \bibnamefont{and} \bibinfo{author}{\bibfnamefont{Y.}~\bibnamefont{Zhang}}, \bibinfo{journal}{Eur. Phys. J. C} \textbf{\bibinfo{volume}{85}}, \bibinfo{pages}{559} (\bibinfo{year}{2025}), \eprint{2412.04881}.

\bibitem[{\citenamefont{P{\^\i}rvu et~al.}(2024)\citenamefont{P{\^\i}rvu, Johnson, and Sibiryakov}}]{Pirvu:2023plk}
\bibinfo{author}{\bibfnamefont{D.}~\bibnamefont{P{\^\i}rvu}}, \bibinfo{author}{\bibfnamefont{M.~C.} \bibnamefont{Johnson}}, \bibnamefont{and} \bibinfo{author}{\bibfnamefont{S.}~\bibnamefont{Sibiryakov}}, \bibinfo{journal}{JHEP} \textbf{\bibinfo{volume}{11}}, \bibinfo{pages}{064} (\bibinfo{year}{2024}), \eprint{2312.13364}.

\bibitem[{\citenamefont{Figueroa et~al.}(2021)\citenamefont{Figueroa, Florio, Torrenti, and Valkenburg}}]{Figueroa:2020rrl}
\bibinfo{author}{\bibfnamefont{D.~G.} \bibnamefont{Figueroa}}, \bibinfo{author}{\bibfnamefont{A.}~\bibnamefont{Florio}}, \bibinfo{author}{\bibfnamefont{F.}~\bibnamefont{Torrenti}}, \bibnamefont{and} \bibinfo{author}{\bibfnamefont{W.}~\bibnamefont{Valkenburg}}, \bibinfo{journal}{JCAP} \textbf{\bibinfo{volume}{04}}, \bibinfo{pages}{035} (\bibinfo{year}{2021}), \eprint{2006.15122}.

\bibitem[{\citenamefont{Figueroa et~al.}(2023)\citenamefont{Figueroa, Florio, Torrenti, and Valkenburg}}]{Figueroa:2021yhd}
\bibinfo{author}{\bibfnamefont{D.~G.} \bibnamefont{Figueroa}}, \bibinfo{author}{\bibfnamefont{A.}~\bibnamefont{Florio}}, \bibinfo{author}{\bibfnamefont{F.}~\bibnamefont{Torrenti}}, \bibnamefont{and} \bibinfo{author}{\bibfnamefont{W.}~\bibnamefont{Valkenburg}}, \bibinfo{journal}{Comput. Phys. Commun.} \textbf{\bibinfo{volume}{283}}, \bibinfo{pages}{108586} (\bibinfo{year}{2023}), \eprint{2102.01031}.

\bibitem[{\citenamefont{Hu et~al.}(2019)\citenamefont{Hu, Li, Anderson, Ragan-Kelley, and Durand}}]{hu2019taichi}
\bibinfo{author}{\bibfnamefont{Y.}~\bibnamefont{Hu}}, \bibinfo{author}{\bibfnamefont{T.-M.} \bibnamefont{Li}}, \bibinfo{author}{\bibfnamefont{L.}~\bibnamefont{Anderson}}, \bibinfo{author}{\bibfnamefont{J.}~\bibnamefont{Ragan-Kelley}}, \bibnamefont{and} \bibinfo{author}{\bibfnamefont{F.}~\bibnamefont{Durand}}, \bibinfo{journal}{ACM Transactions on Graphics (TOG)} \textbf{\bibinfo{volume}{38}}, \bibinfo{pages}{1} (\bibinfo{year}{2019}).

\bibitem[{\citenamefont{Dankovsky et~al.}(2024)\citenamefont{Dankovsky, Babichev, Gorbunov, Ramazanov, and Vikman}}]{Dankovsky:2024zvs}
\bibinfo{author}{\bibfnamefont{I.}~\bibnamefont{Dankovsky}}, \bibinfo{author}{\bibfnamefont{E.}~\bibnamefont{Babichev}}, \bibinfo{author}{\bibfnamefont{D.}~\bibnamefont{Gorbunov}}, \bibinfo{author}{\bibfnamefont{S.}~\bibnamefont{Ramazanov}}, \bibnamefont{and} \bibinfo{author}{\bibfnamefont{A.}~\bibnamefont{Vikman}}, \bibinfo{journal}{JCAP} \textbf{\bibinfo{volume}{09}}, \bibinfo{pages}{047} (\bibinfo{year}{2024}), \eprint{2406.17053}.

\bibitem[{\citenamefont{Cutting et~al.}(2018)\citenamefont{Cutting, Hindmarsh, and Weir}}]{Cutting:2018tjt}
\bibinfo{author}{\bibfnamefont{D.}~\bibnamefont{Cutting}}, \bibinfo{author}{\bibfnamefont{M.}~\bibnamefont{Hindmarsh}}, \bibnamefont{and} \bibinfo{author}{\bibfnamefont{D.~J.} \bibnamefont{Weir}}, \bibinfo{journal}{Phys. Rev. D} \textbf{\bibinfo{volume}{97}}, \bibinfo{pages}{123513} (\bibinfo{year}{2018}), \eprint{1802.05712}.

\bibitem[{\citenamefont{Cutting et~al.}(2021)\citenamefont{Cutting, Escartin, Hindmarsh, and Weir}}]{Cutting:2020nla}
\bibinfo{author}{\bibfnamefont{D.}~\bibnamefont{Cutting}}, \bibinfo{author}{\bibfnamefont{E.~G.} \bibnamefont{Escartin}}, \bibinfo{author}{\bibfnamefont{M.}~\bibnamefont{Hindmarsh}}, \bibnamefont{and} \bibinfo{author}{\bibfnamefont{D.~J.} \bibnamefont{Weir}}, \bibinfo{journal}{Phys. Rev. D} \textbf{\bibinfo{volume}{103}}, \bibinfo{pages}{023531} (\bibinfo{year}{2021}), \eprint{2005.13537}.

\bibitem[{\citenamefont{Schmitz}(2021)}]{Schmitz:2020syl}
\bibinfo{author}{\bibfnamefont{K.}~\bibnamefont{Schmitz}}, \bibinfo{journal}{JHEP} \textbf{\bibinfo{volume}{01}}, \bibinfo{pages}{097} (\bibinfo{year}{2021}), \eprint{2002.04615}.

\bibitem[{\citenamefont{Hindmarsh and Hijazi}(2019)}]{Hindmarsh:2019phv}
\bibinfo{author}{\bibfnamefont{M.}~\bibnamefont{Hindmarsh}} \bibnamefont{and} \bibinfo{author}{\bibfnamefont{M.}~\bibnamefont{Hijazi}}, \bibinfo{journal}{JCAP} \textbf{\bibinfo{volume}{12}}, \bibinfo{pages}{062} (\bibinfo{year}{2019}), \eprint{1909.10040}.

\bibitem[{\citenamefont{Harry et~al.}(2006)\citenamefont{Harry, Fritschel, Shaddock, Folkner, and Phinney}}]{Harry_2006}
\bibinfo{author}{\bibfnamefont{G.~M.} \bibnamefont{Harry}}, \bibinfo{author}{\bibfnamefont{P.}~\bibnamefont{Fritschel}}, \bibinfo{author}{\bibfnamefont{D.~A.} \bibnamefont{Shaddock}}, \bibinfo{author}{\bibfnamefont{W.}~\bibnamefont{Folkner}}, \bibnamefont{and} \bibinfo{author}{\bibfnamefont{E.~S.} \bibnamefont{Phinney}}, \bibinfo{journal}{Classical and Quantum Gravity} \textbf{\bibinfo{volume}{23}}, \bibinfo{pages}{4887} (\bibinfo{year}{2006}), \urlprefix\url{https://doi.org/10.1088/0264-9381/23/15/008}.

\bibitem[{\citenamefont{Corbin and Cornish}(2006)}]{Corbin_2006}
\bibinfo{author}{\bibfnamefont{V.}~\bibnamefont{Corbin}} \bibnamefont{and} \bibinfo{author}{\bibfnamefont{N.~J.} \bibnamefont{Cornish}}, \bibinfo{journal}{Classical and Quantum Gravity} \textbf{\bibinfo{volume}{23}}, \bibinfo{pages}{2435} (\bibinfo{year}{2006}), \urlprefix\url{https://doi.org/10.1088/0264-9381/23/7/014}.

\bibitem[{\citenamefont{Crowder and Cornish}(2005)}]{PhysRevD.72.083005}
\bibinfo{author}{\bibfnamefont{J.}~\bibnamefont{Crowder}} \bibnamefont{and} \bibinfo{author}{\bibfnamefont{N.~J.} \bibnamefont{Cornish}}, \bibinfo{journal}{Phys. Rev. D} \textbf{\bibinfo{volume}{72}}, \bibinfo{pages}{083005} (\bibinfo{year}{2005}), \urlprefix\url{https://link.aps.org/doi/10.1103/PhysRevD.72.083005}.

\bibitem[{\citenamefont{Isoyama et~al.}(2018)\citenamefont{Isoyama, Nakano, and Nakamura}}]{10.1093/ptep/pty078}
\bibinfo{author}{\bibfnamefont{S.}~\bibnamefont{Isoyama}}, \bibinfo{author}{\bibfnamefont{H.}~\bibnamefont{Nakano}}, \bibnamefont{and} \bibinfo{author}{\bibfnamefont{T.}~\bibnamefont{Nakamura}}, \bibinfo{journal}{Progress of Theoretical and Experimental Physics} \textbf{\bibinfo{volume}{2018}}, \bibinfo{pages}{073E01} (\bibinfo{year}{2018}), ISSN \bibinfo{issn}{2050-3911}, \eprint{https://academic.oup.com/ptep/article-pdf/2018/7/073E01/25332865/pty078.pdf}, \urlprefix\url{https://doi.org/10.1093/ptep/pty078}.

\bibitem[{\citenamefont{Yagi and Seto}(2011)}]{PhysRevD.83.044011}
\bibinfo{author}{\bibfnamefont{K.}~\bibnamefont{Yagi}} \bibnamefont{and} \bibinfo{author}{\bibfnamefont{N.}~\bibnamefont{Seto}}, \bibinfo{journal}{Phys. Rev. D} \textbf{\bibinfo{volume}{83}}, \bibinfo{pages}{044011} (\bibinfo{year}{2011}), \urlprefix\url{https://link.aps.org/doi/10.1103/PhysRevD.83.044011}.

\bibitem[{\citenamefont{Seto et~al.}(2001)\citenamefont{Seto, Kawamura, and Nakamura}}]{PhysRevLett.87.221103}
\bibinfo{author}{\bibfnamefont{N.}~\bibnamefont{Seto}}, \bibinfo{author}{\bibfnamefont{S.}~\bibnamefont{Kawamura}}, \bibnamefont{and} \bibinfo{author}{\bibfnamefont{T.}~\bibnamefont{Nakamura}}, \bibinfo{journal}{Phys. Rev. Lett.} \textbf{\bibinfo{volume}{87}}, \bibinfo{pages}{221103} (\bibinfo{year}{2001}), \urlprefix\url{https://link.aps.org/doi/10.1103/PhysRevLett.87.221103}.

\bibitem[{\citenamefont{Amaro-Seoane et~al.}(2017)\citenamefont{Amaro-Seoane, Audley, Babak, Baker, Barausse, Bender, Berti, Binetruy, Born, Bortoluzzi et~al.}}]{amaro2017laser}
\bibinfo{author}{\bibfnamefont{P.}~\bibnamefont{Amaro-Seoane}}, \bibinfo{author}{\bibfnamefont{H.}~\bibnamefont{Audley}}, \bibinfo{author}{\bibfnamefont{S.}~\bibnamefont{Babak}}, \bibinfo{author}{\bibfnamefont{J.}~\bibnamefont{Baker}}, \bibinfo{author}{\bibfnamefont{E.}~\bibnamefont{Barausse}}, \bibinfo{author}{\bibfnamefont{P.}~\bibnamefont{Bender}}, \bibinfo{author}{\bibfnamefont{E.}~\bibnamefont{Berti}}, \bibinfo{author}{\bibfnamefont{P.}~\bibnamefont{Binetruy}}, \bibinfo{author}{\bibfnamefont{M.}~\bibnamefont{Born}}, \bibinfo{author}{\bibfnamefont{D.}~\bibnamefont{Bortoluzzi}}, \bibnamefont{et~al.}, \bibinfo{journal}{arXiv preprint arXiv:1702.00786}  (\bibinfo{year}{2017}).

\bibitem[{\citenamefont{Baker et~al.}(2019)}]{Baker:2019nia}
\bibinfo{author}{\bibfnamefont{J.}~\bibnamefont{Baker}} \bibnamefont{et~al.} (\bibinfo{year}{2019}), \eprint{1907.06482}.

\bibitem[{\citenamefont{Ruan et~al.}(2020)\citenamefont{Ruan, Guo, Cai, and Zhang}}]{taiji}
\bibinfo{author}{\bibfnamefont{W.-H.} \bibnamefont{Ruan}}, \bibinfo{author}{\bibfnamefont{Z.-K.} \bibnamefont{Guo}}, \bibinfo{author}{\bibfnamefont{R.-G.} \bibnamefont{Cai}}, \bibnamefont{and} \bibinfo{author}{\bibfnamefont{Y.-Z.} \bibnamefont{Zhang}}, \bibinfo{journal}{International Journal of Modern Physics A} \textbf{\bibinfo{volume}{35}}, \bibinfo{pages}{2050075} (\bibinfo{year}{2020}), \eprint{https://doi.org/10.1142/S0217751X2050075X}, \urlprefix\url{https://doi.org/10.1142/S0217751X2050075X}.

\bibitem[{\citenamefont{Hu and Wu}(2017)}]{taijii}
\bibinfo{author}{\bibfnamefont{W.-R.} \bibnamefont{Hu}} \bibnamefont{and} \bibinfo{author}{\bibfnamefont{Y.-L.} \bibnamefont{Wu}}, \bibinfo{journal}{National Science Review} \textbf{\bibinfo{volume}{4}}, \bibinfo{pages}{685} (\bibinfo{year}{2017}), ISSN \bibinfo{issn}{2095-5138}, \eprint{https://academic.oup.com/nsr/article-pdf/4/5/685/31566708/nwx116.pdf}, \urlprefix\url{https://doi.org/10.1093/nsr/nwx116}.

\bibitem[{\citenamefont{Luo et~al.}(2016)\citenamefont{Luo, Chen, Duan, Gong, Hu, Ji, Liu, Mei, Milyukov, Sazhin et~al.}}]{Luo_2016}
\bibinfo{author}{\bibfnamefont{J.}~\bibnamefont{Luo}}, \bibinfo{author}{\bibfnamefont{L.-S.} \bibnamefont{Chen}}, \bibinfo{author}{\bibfnamefont{H.-Z.} \bibnamefont{Duan}}, \bibinfo{author}{\bibfnamefont{Y.-G.} \bibnamefont{Gong}}, \bibinfo{author}{\bibfnamefont{S.}~\bibnamefont{Hu}}, \bibinfo{author}{\bibfnamefont{J.}~\bibnamefont{Ji}}, \bibinfo{author}{\bibfnamefont{Q.}~\bibnamefont{Liu}}, \bibinfo{author}{\bibfnamefont{J.}~\bibnamefont{Mei}}, \bibinfo{author}{\bibfnamefont{V.}~\bibnamefont{Milyukov}}, \bibinfo{author}{\bibfnamefont{M.}~\bibnamefont{Sazhin}}, \bibnamefont{et~al.}, \bibinfo{journal}{Classical and Quantum Gravity} \textbf{\bibinfo{volume}{33}}, \bibinfo{pages}{035010} (\bibinfo{year}{2016}), \urlprefix\url{https://doi.org/10.1088/0264-9381/33/3/035010}.

\bibitem[{\citenamefont{Zhou et~al.}(2023)\citenamefont{Zhou, Cheng, and Ren}}]{Zhou:2023rop}
\bibinfo{author}{\bibfnamefont{K.}~\bibnamefont{Zhou}}, \bibinfo{author}{\bibfnamefont{J.}~\bibnamefont{Cheng}}, \bibnamefont{and} \bibinfo{author}{\bibfnamefont{L.}~\bibnamefont{Ren}} (\bibinfo{year}{2023}), \eprint{2306.14439}.

\end{thebibliography}

\clearpage
\begin{widetext}
\setcounter{equation}{0}
\renewcommand{\theequation}{S\arabic{equation}}
\setcounter{figure}{0}
\renewcommand{\thefigure}{S\arabic{figure}}
\setcounter{table}{0}
\renewcommand{\thetable}{S\arabic{table}}
\section*{Supplemental Material}
This supplemental material provides details of our simulations and further explanations of the results presented in the main text, along with additional supporting results.

\section*{Potential and Bubble Profile}
We consider the finite-temperature effective potential
\begin{align}
V(\phi,T)
=&\ \frac{1}{2}\gamma\left(T^{2}-T_{0}^{2}\right)\phi^{2}
-\frac{1}{3}\alpha\,T\,\phi^{3}
+\frac{1}{4}\lambda\,\phi^{4} \notag\\
\equiv&\ \frac{1}{2}M^{2}(T)\phi^{2}
-\frac{1}{3}\delta(T)\phi^{3}
+\frac{1}{4}\lambda\,\phi^{4},
\end{align}
where $M^{2}(T)\equiv \gamma\left(T^{2}-T_{0}^{2}\right)$ and $\delta(T)\equiv \alpha T$. For our benchmark parameter choice $(\gamma,\alpha,\lambda)=(1.11,\,1.41,\,0.5)$, the potential exhibits two minimums: the symmetric-phase vacuum $\phi_s$ and the broken-phase vacuum
\begin{equation}
\phi_b=\frac{\delta+\sqrt{\delta^{2}-4M^{2}\lambda}}{2\lambda}.
\end{equation}
At $T=50~\mathrm{GeV},\,55~\mathrm{GeV},\,60~\mathrm{GeV}$, it gives
\begin{equation}
\phi_b = 108.15~\mathrm{GeV},\quad 113.55~\mathrm{GeV},\quad 118.70~\mathrm{GeV},
\end{equation}
respectively. The schematic of the potential is shown in the left panel of Fig.~\ref{fig: potential and profile}. The critical bubble profile can be obtained as a stationary point of the Euclidean action. In the one-dimensional case, we solve it using the shooting method, and the result is shown in the right panel of Fig.~\ref{fig: potential and profile}. We find that as $T$ increases, the critical profile becomes closer to a $\tanh$-like shape, whereas for smaller $T$ it is better approximated by a Gaussian-like form.
\begin{figure}[htbp!]
\centering 
\includegraphics[width=0.45\textwidth]{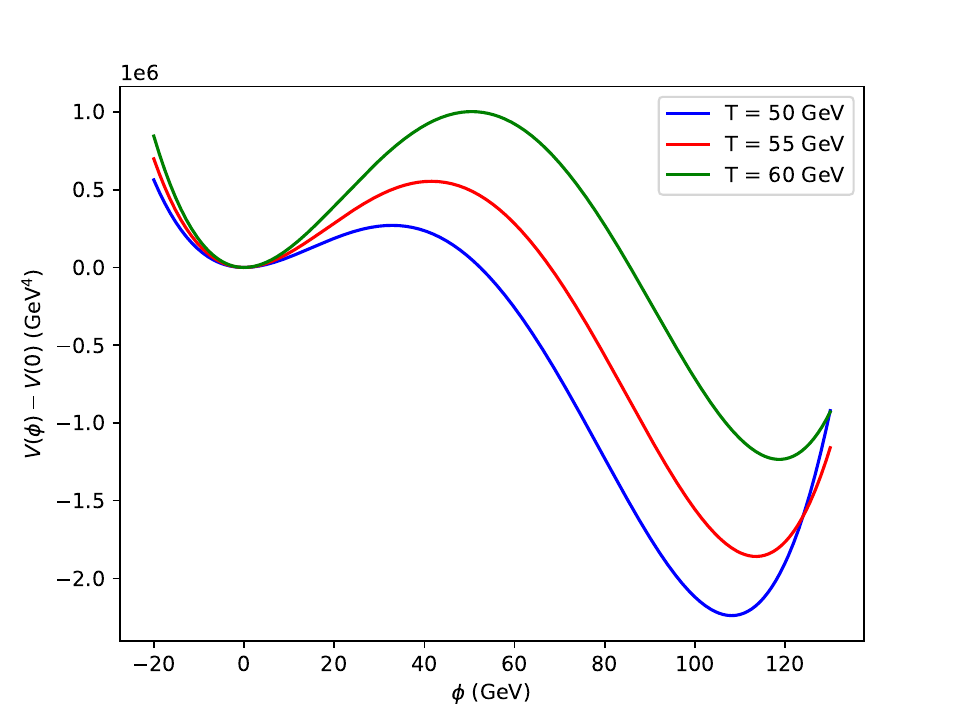}
\includegraphics[width=0.45\textwidth]{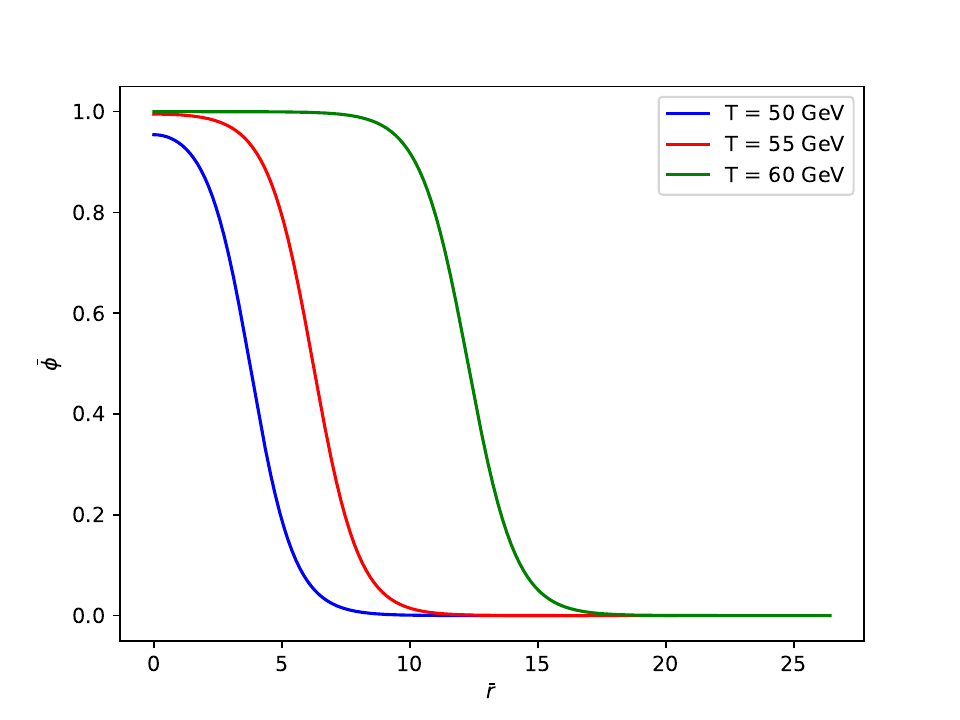}
\caption{(Left): Schematic illustration of the scalar potential. (Right): Schematic illustration of the critical bubble profile.}
\label{fig: potential and profile}
\end{figure}

\section*{Thermal Fluctuation and Symmetry Factor $\epsilon$}
In practice, the definition of $\epsilon$ used in our simulations does not strictly follow Eq.~(5). This is because, when thermal fluctuations are present, the maximum and minimum values of the scalar field change with time.
As a result, for a fixed threshold $\alpha$, a surface of constant field value may not exist in all directions, and the corresponding radius $R_{\alpha}$ can become ill-defined.

To avoid this problem, we normalize the scalar field at each time step as
\begin{equation}
    \tilde{\phi}(\bar t)
    =
    \frac{\bar{\phi}(\bar t)-\bar{\phi}_{\rm min}(\bar t)}
         {\bar{\phi}_{\rm max}(\bar t)-\bar{\phi}_{\rm min}(\bar t)},
\end{equation}
which maps the field values to the range $[0,1]$.
We then define the radius $R_{\alpha}$ through the condition
$\tilde{\phi}(R_{\alpha},\bar t)=\alpha$.
With this definition, the symmetry parameter
\begin{equation}
    \epsilon =  \frac{{\rm Std}[R(\hat{\mathbf n},\bar t)]}{{\rm Avg}[R(\hat{\mathbf n},\bar t)]}
\end{equation}
can be evaluated in a stable way at all times. We also note that this definition of $\epsilon$ does not capture the strong field oscillations inside the bubble,
but defining a measure of spherical symmetry in the bubble interior is beyond the scope of the present study.

In our simulation, the thermal fluctuations are sampled through the following equations
\begin{align} 
    \bra{0}\delta\phi(\bold{k})\delta \phi^{*}(\bold{k'})\ket{0}  &= \frac{(2\pi)^3\delta^{3}(\bold{k}-\bold{k'})}{w_k} \frac{1}{e^{\frac{w_k}{T}}-1},\\
\bra{0}\delta \pi(\bold{k}) \delta \pi^{*}(\bold{k'})\ket{0} &= (2\pi)^3\delta^{3}(\bold{k}-\bold{k'})  \frac{w_k}{e^{\frac{w_k}{T}}-1}, \\
\bra{0}\delta\phi(\bold{k})\delta \pi^{*}(\bold{k'})\ket{0} &= 0,
\end{align}
We can approximate this distribution by a Gaussian random distribution with zero mean and variance $\delta$, where $\delta$ is computed from the above expression. After sampling, we apply the discrete inverse Fourier transform to obtain the thermal fluctuations in position space. We show the initial thermal fluctuation field distribution in Fig.~\ref{fig: f_dis}. Because the potential barrier is relatively high at large temperatures, only a small fraction of thermal fluctuations can overcome it. Such rare excursions can locally seed the true vacuum. However, since the affected regions are not large enough, they cannot expand and instead undergo oscillatory motion. As the temperature decreases, although the fluctuation amplitude becomes smaller, the potential barrier also decreases. In our model, the latter effect is more pronounced, so compared to the high-temperature case, more thermal fluctuations can overcome the barrier. This is why, in Fig.~2, the symmetry factor exhibits larger perturbations at lower temperatures.

\begin{figure}[htbp!]
\centering 
\includegraphics[width=0.45\textwidth]{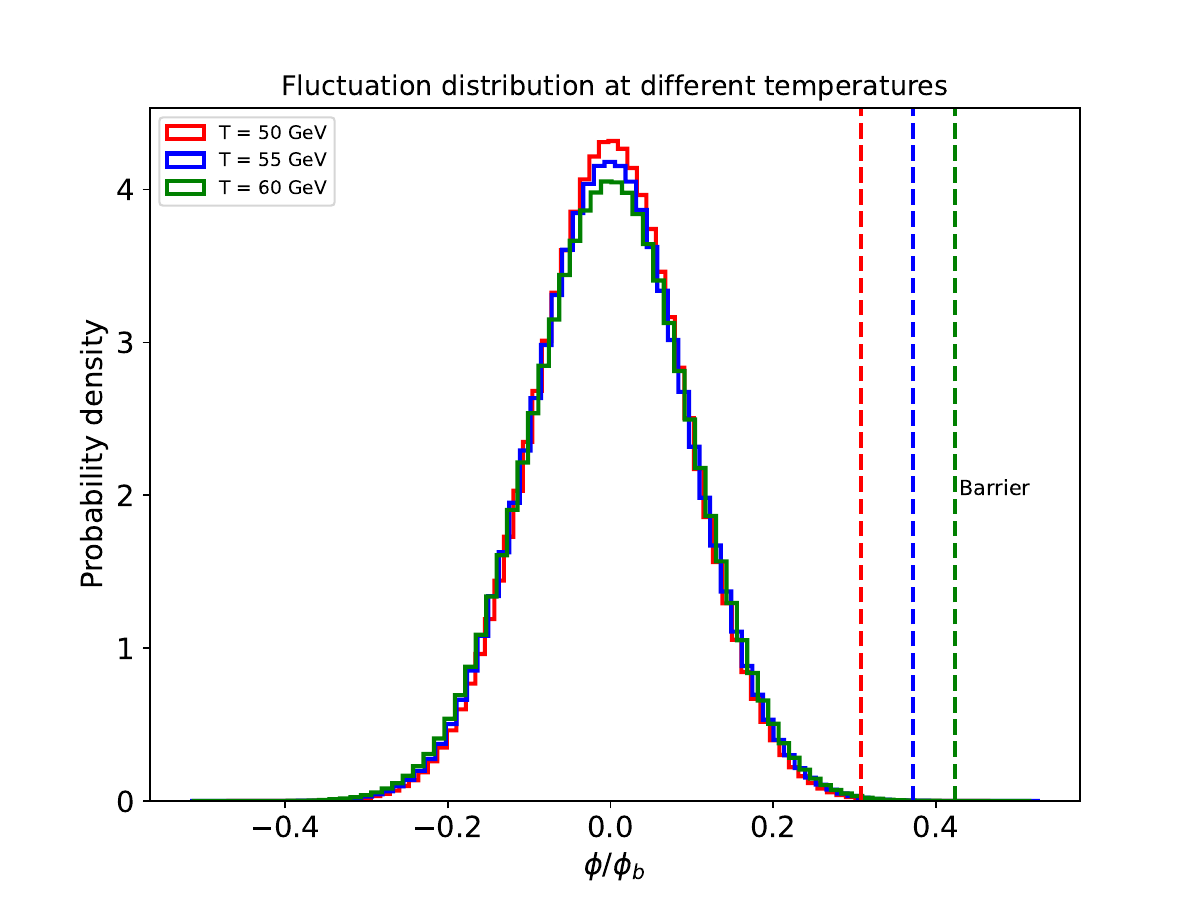}
\caption{The sampled density distributions of the scalar field $\bar{\phi}$ at different temperatures, with the dashed line indicating the position of the potential barrier.}
\label{fig: f_dis}
\end{figure}

\section*{Gravitational Wave Spectrum}
To compute GW spectrum, we first solve the scalar-field equation of motion with specified initial conditions,
\begin{equation}
    \frac{\partial^2 \phi}{\partial t^2}-\nabla^2\phi
    = -\frac{\partial V(\phi)}{\partial \phi}\, .
\end{equation}
Once the time evolution of $\phi$ is obtained, we construct the corresponding energy--momentum tensor. Since only the anisotropic stress sources GWs, we retain the spatial, traceless contribution and write
\begin{equation}
    T_{ij} = \partial_i\phi\,\partial_j\phi \, ,
\end{equation}
where we have omitted terms that do not contribute to the transverse--traceless (TT) source.

The GW field $h_{ij}$ then follows from the linearized Einstein equation,
\begin{equation}
    \ddot{h}_{ij}-\nabla^2 h_{ij}
    = 16\pi G\,\Lambda_{ij,kl}\,T_{kl}\, ,
\end{equation}
where $\Lambda_{ij,kl}$ denotes the TT projection operator. The GW energy density is defined as
\begin{equation}
    \rho_{\rm GW}
    = \frac{1}{32\pi G}\left\langle \dot{h}_{ij}(\mathbf{x},t)\,\dot{h}_{ij}(\mathbf{x},t)\right\rangle
    = \frac{1}{32\pi G\,L^3}\int d^3\mathbf{x}\; \dot{h}_{ij}(\mathbf{x},t)\,\dot{h}_{ij}(\mathbf{x},t)\, .
\end{equation}
Accordingly, the GW spectrum per logarithmic wavenumber interval is
\begin{align}
    \Omega_{\rm GW} &= \int \frac{dk}{k}\,\frac{d\Omega_{\rm GW}}{d\ln k}\, ,\\
    \frac{d\Omega_{\rm GW}}{d\ln k}
    &= \frac{1}{\rho_c}\frac{d\rho_{\rm GW}}{d\ln k}
    = \frac{k^3}{32\pi G\,\rho_c\,L^3}\int d\Omega\;
    \dot{h}_{ij}(\mathbf{k},t)\,\dot{h}^{*}_{ij}(\mathbf{k},t)\, .
\end{align}

To verify that our lattice GW results are reliable in the ultraviolet regime, we focus on the wavenumber range around the second peak and vary the lattice spacing $d\bar{x}$ ($d\bar{x}=0.11,\ 0.22,\ 0.44$). The results are shown in Fig~\ref{fig: Convergence tests}.
\begin{figure}[htbp!]
\centering 
\includegraphics[width=0.45\textwidth]{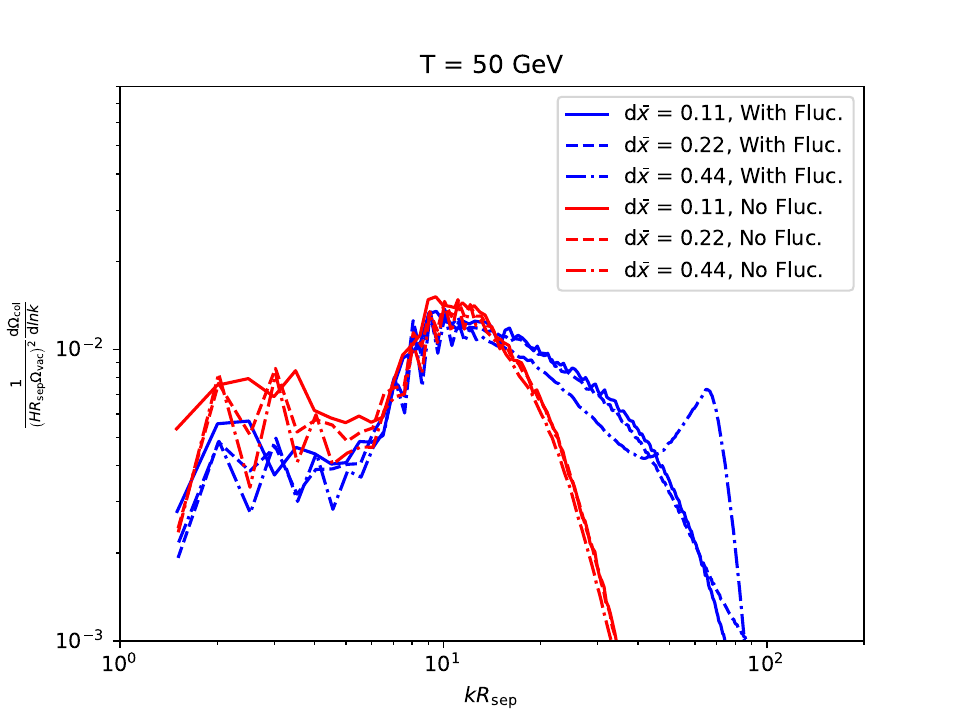}
\includegraphics[width=0.45\textwidth]{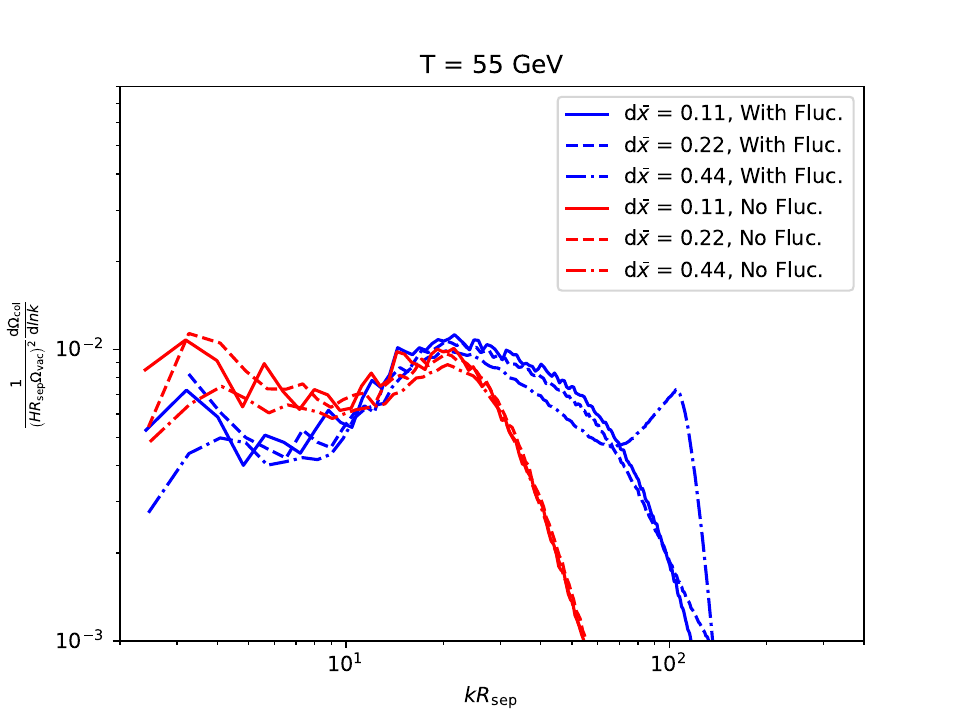}
\includegraphics[width=0.45\textwidth]{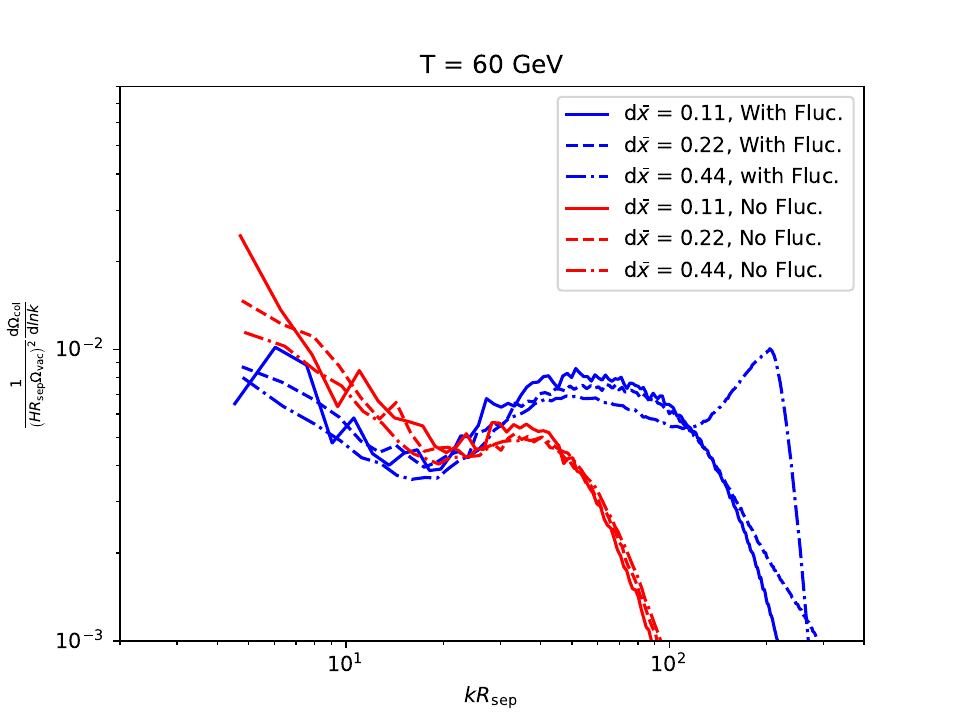}
\caption{Convergence tests of the lattice GW results.}
\label{fig: Convergence tests}
\end{figure}

We find that an ultraviolet peak is present for all three lattice spacings, confirming that thermal fluctuations robustly enhance the spectrum at high wavenumber. However, the run with $d\bar{x}=0.44$ yields systematically lower power than the finer resolutions and exhibits an additional spurious peak, which we attribute to an algorithm-induced numerical artifact. This feature disappears upon lattice refinement. For $T=50~\mathrm{GeV}$ and $55~\mathrm{GeV}$, the results obtained with $d\bar{x}=0.11$ and $d\bar{x}=0.22$ agree well, whereas for $T=60~\mathrm{GeV}$ the $d\bar{x}=0.11$ result is slightly larger than the $d\bar{x}=0.22$ one. Although thermal fluctuations somewhat weaken ultraviolet convergence compared to the fluctuation-free case, the spectra remain well converged over the wavenumber range relevant to our analysis. We therefore conclude that $d\bar{x}=0.44$ is inadequate for a quantitative determination of the ultraviolet spectrum, while $d\bar{x}=0.22$ provides reliable ultraviolet results.

\end{widetext}

\end{document}